\documentclass[a4paper]{article}

\usepackage{url}

\usepackage{graphicx}
\usepackage{enumitem}
\usepackage{algorithm}
\usepackage{algorithmicx}
\usepackage{algpseudocode}
\usepackage{multirow}
\usepackage{float}
\usepackage{bm}
\usepackage{graphicx}
\usepackage{multicol}
\usepackage{booktabs}
\usepackage[export]{adjustbox}
\usepackage{algpseudocode,amsfonts,threeparttable,booktabs}
\usepackage{amsmath}
\usepackage{pifont}
\usepackage{subfigure}
\usepackage{mathrsfs}
\usepackage{verbatim}
\usepackage{siunitx}
\usepackage{cite}
\usepackage{siunitx}
\usepackage{url}

\usepackage{array}
\usepackage{spconf,amsmath,graphicx}
\usepackage{multirow}
\usepackage{stfloats}
\usepackage{color}
\usepackage{booktabs}
\usepackage[colorlinks,linkcolor=blue]{hyperref}

\title{A Unified Deep Speaker Embedding Framework for Mixed-Bandwidth Speech Data}
\name{Weicheng Cai$^{1,3}$ \qquad Ming Li$^{1,2}$}

\address{$^{1}$Data Science Research Center, Duke Kunshan University, Kunshan, Chain\\
  $^{2}$School of Computer Science, Wuhan University, Wuhan, China  \\
      $^{3}$School of Electronics and Information Technology, Sun Yat-sen University, Guangzhou, China}

\begin{document}
\ninept

\maketitle

\begin{abstract}
This paper proposes a unified deep speaker embedding framework for modeling speech data with different sampling rates. Considering the narrowband spectrogram as a sub-image of the wideband spectrogram, we tackle the joint modeling problem of the mixed-bandwidth data in an image classification manner. From this perspective, we elaborate several mixed-bandwidth joint training strategies under different training and test data scenarios. The proposed systems are able to flexibly handle the mixed-bandwidth speech data in a single speaker embedding model without any additional downsampling, upsampling, bandwidth extension, or padding operations. We conduct extensive experimental studies on the VoxCeleb1 dataset. Furthermore, the effectiveness of the proposed approach is validated by the SITW and NIST SRE 2016 datasets.
\end{abstract}

\begin{keywords}
mixed-bandwidth, unified model, speaker embedding, image classification, convolutional neural network
\end{keywords}


\section{Introduction}

A speech signal can be considered as a variable-length temporal sequence~\cite{graves2012supervised}, and many features have been used to characterize its pattern. Short-term spectral features are used extensively because of the quasi-stationary property of the speech signal. After short-term processing, the raw waveform is converted into a two-dimensional~(2-D) matrix of size $D \times T$, where $D$ represents the frequential feature dimension related to the number of filter coefficients, and $T$ denotes the temporal frame length related to the utterance duration.

For a text-independent speaker verification~(TISV) system, the main procedure is to extract the fixed-dimensional speaker representation from the variable-length spectral feature sequence. One of the widely used spectral features is the Mel-frequency cepstral coefficient ~(MFCC)~\cite{davis1980comparison, Kinnunen2010An}. Typically, MFCC feature vectors from all the frames are assumed to be independent and identically distributed. They can be projected on the Gaussian components or phonetic units to accumulate statistics over the time axis and form a high-dimensional supervector. Then, a factor analysis-based dimension reduction is performed to generate a fixed-dimensional low rank i-vector representation~\cite{dehak2010front}. Recently, with the progress of deep learning, many approaches directly train a deep neural network~(DNN) to distinguish different speakers~\cite{variani2014deep, 2017arXiv170502304L, Snyder2017Deep, xvector, Nagrani2017, Cai_2018_Odyssey}. Systems comprising of x-vector~\cite{xvector} speaker embedding followed by a probabilistic linear discriminant analysis~(PLDA)~\cite{prince2007probabilistic} have shown state-of-the-art performances on multiple TISV tasks~\cite{xvector}. In the x-vector system, a time-delay neural network~(TDNN)~\cite{21701} followed by statistic pooling over the time axis is used for modeling the long-term temporal dependencies from the MFCC features.

\begin{figure*}
\setlength{\abovecaptionskip}{-0.3cm}
    \centering
    \begin{multicols}{3}
    \setlength{\abovecaptionskip}{0.cm}
        \subfigure[\label{fig:plot1}]{\includegraphics[width=0.27\textwidth]{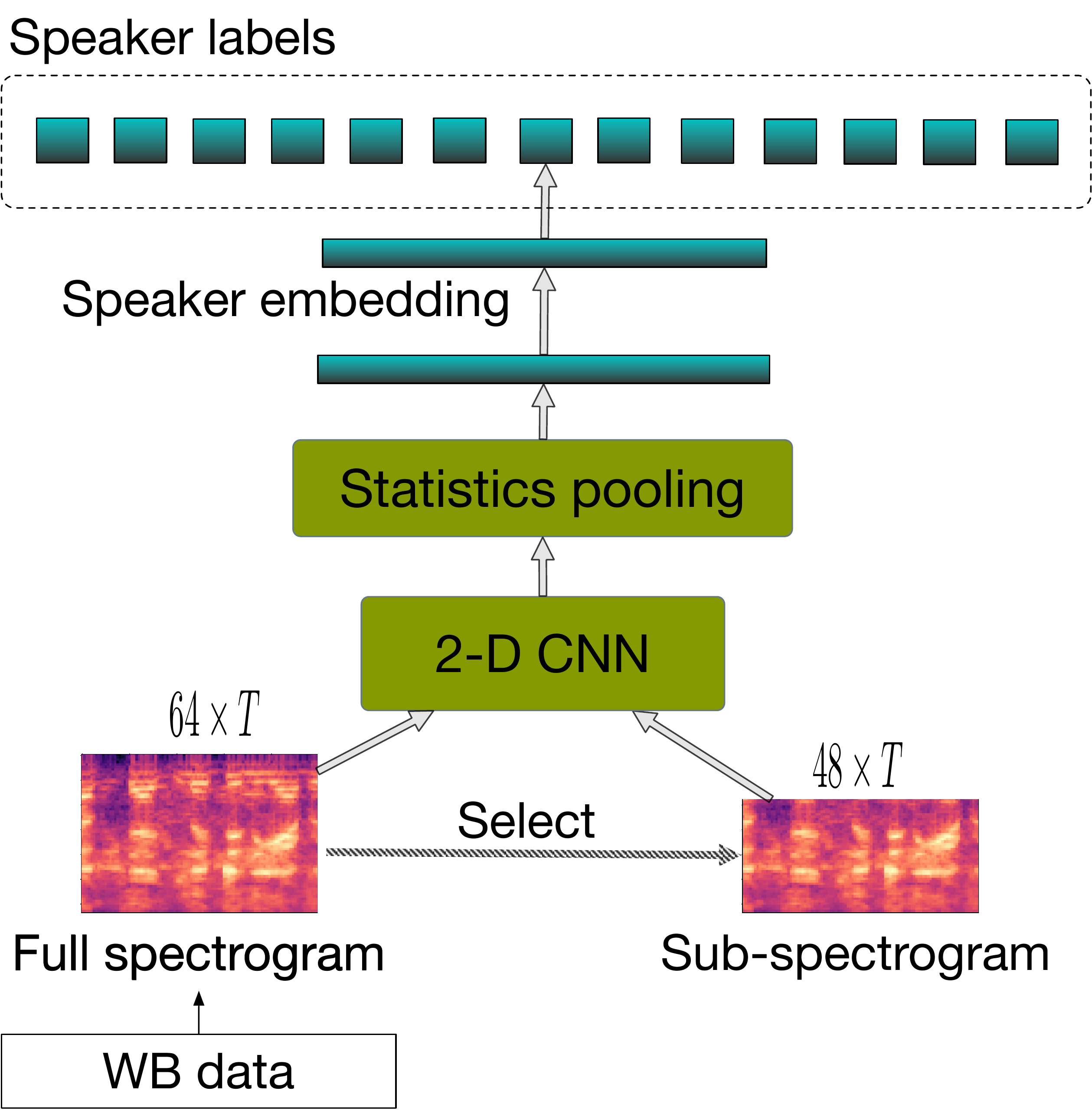}\par }
        \subfigure[\label{fig:plot2}]{\includegraphics[width=0.32\textwidth]{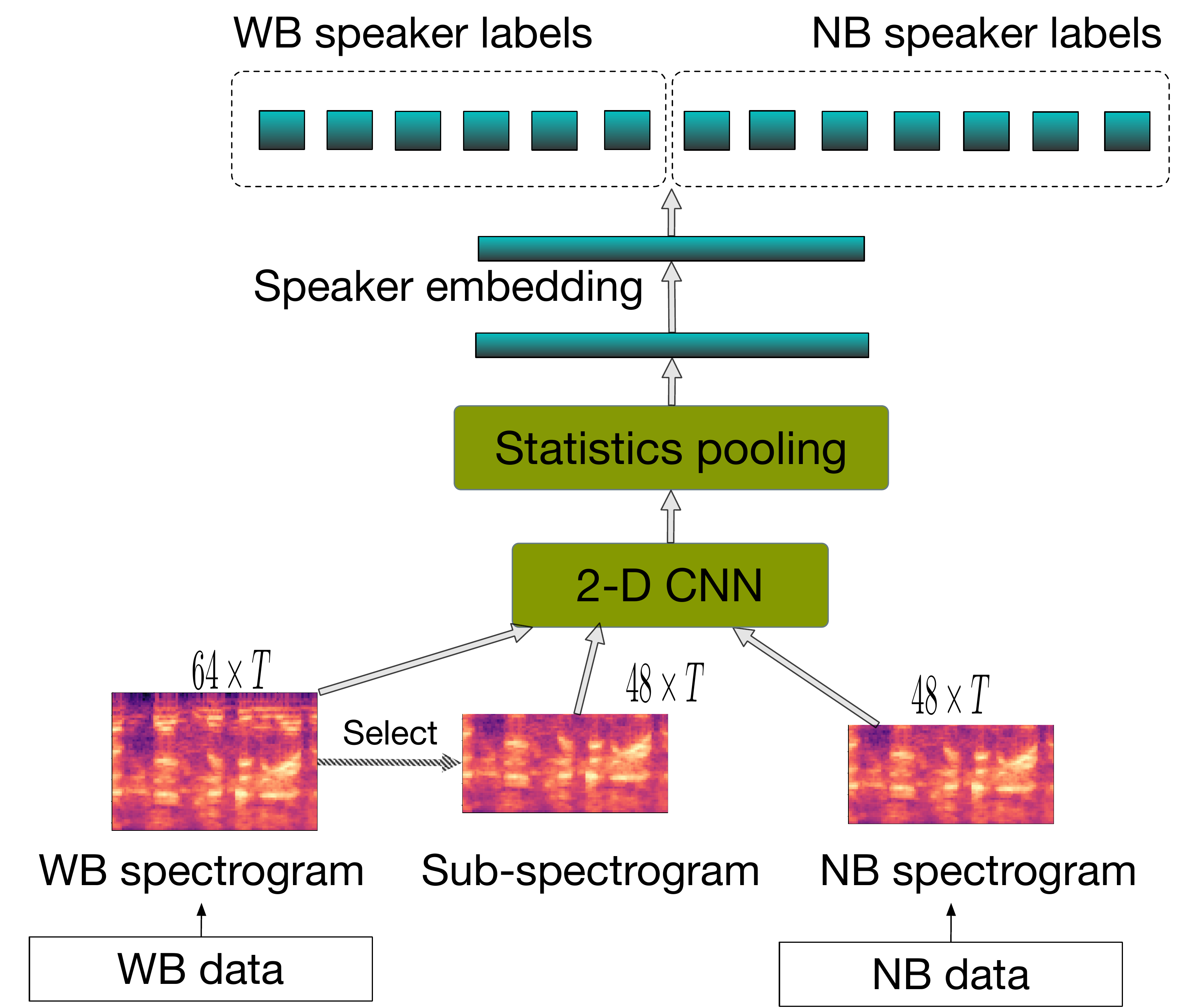} \par }
        \subfigure[\label{fig:plot3}]{\includegraphics[width=0.37\textwidth]{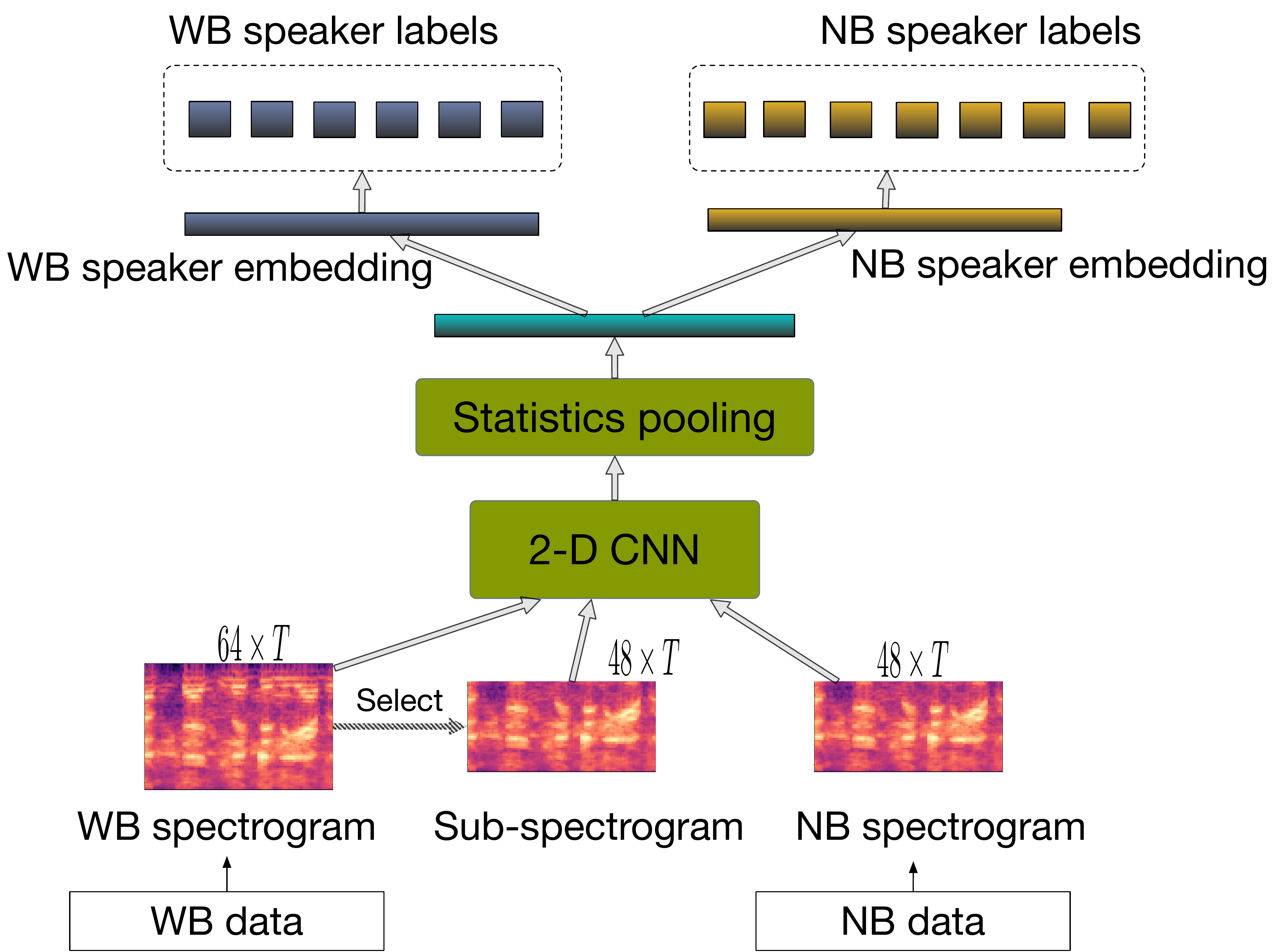}\par}
 \end{multicols}
    \caption{Three mixed-bandwidth data training strategies for different scenarios: (a) Only WB training data are given; (b) WB and NB training data are from the same domain; (c) WB and NB training data are from different domains }
    \label{fig:losscurve}
\vspace{-0.5cm}
\end{figure*}

For the i-vector, x-vector, and many other speech modeling methods, the feature matrix $D \times T$ is viewed as a multi-channel 1-D time series. Although the duration $T$ may vary among the utterances, the feature dimension $D$ must be a fixed value. In this paper, we consider the feature matrix as a single-channel 2-D image~\cite{lecun1995convolutional}. From this new perspective, the spectral feature is viewed as a ``picture" of the sound, and a 2-D CNN  is implemented in the same way as traditional image recognition paradigms. This kind of process brings a type of flexibility, i.e., the size of the input ``image," including the width (frame length) and the height (feature dimension), can be arbitrary numbers.  In other words, a 2-D CNN trained with a 64-dimensional spectrogram could potentially also process a spectrogram with 48 dimensions.

We aim to utilize the flexibility of the 2-D CNN to tackle the mixed-bandwidth~(MB) joint modeling problem. Currently, there are many devices and equipment that capture speech data in different sampling rates, thus solving the sampling rate mismatch problem has become a research topic in the speech community. The traditional way to accomplish this goal is to train a specific model for every target bandwidth since the sampling rates are different (typically 8k Hz vs. 16k Hz). An alternative solution is to uniformly downsample the wideband~(WB) speech data or extend the bandwidth of a narrowband~(NB) data, so that they can be combined ~\cite{yamamoto2019speaker,yingxuewang2015}.

In this paper, we present a unified solution to solve the MB joint modeling problem. The key idea is to view the NB spectrogram as a sub-image of the WB spectrogram. The major contributions of this work are summarized as follows.

\begin{itemize}[noitemsep]
    \item We leverage the 2-D CNN to tackle the MB joint modeling problem from a novel image classification perspective. We show that speech data with different bandwidths can be naturally combined without any additional downsamping, upsampling, bandwidth extension, padding operation, or auxiliary input.
   \item We further investigate several network training strategies targeting various real-world MB application scenarios, including when (a) only the WB training data are given; (b) WB and NB training data are given from the same domain; and (c) WB and NB training data are given but from different domains.
\end{itemize}

\section{Related works}

In~\cite{4032793}, Seltzer~\textit{et al.} present an expectation-maximization algorithm for training with MB data where the missing spectral components of the NB signal are considered additional hidden variables. Li\textit{ et al.} formulate the MB joint training problem as a missing data paradigm and propose training an MB speech recognition system without bandwidth extension in~\cite{li2012improving}. The authors adopt a fully-connected DNN architecture, and thus require a zero-padding or mean-padding operation to ensure all features have the same dimensions. In~\cite{mantena2019bandwidth}, Gautam~\textit{et al.} build a single model for MB speech recognition. The inputs of their network are fixed 40-dimensional features, and their network requires a bandwidth embedding as the auxiliary input.

Recently, since there are more and more open speech databases with speaker labels collected in different sampling rates, the MB joint modeling problem has gained much attention in the speaker recognition community. In~\cite{nidadavolu2018investigation, 8682992}, Nidadavolu~\textit{et al.} investigated several bandwidth extension approaches for speaker recognition with several different network architectures. Meanwhile, the authors of~\cite{yamamoto2019speaker} consider making use of the Mel filter bank coefficients to share acoustic features between WB and NB speech, and implement a new pipeline that uses a DNN-based bandwidth extension as pre-processing of the DNN for speaker embedding extraction.

\section{Methods}

\subsection{Mel-spectrogram}

We adopt the log Mel-filterbank energies as the standard acoustic features. We refer to this feature as the Mel-spectrogram because we process it in an image processing manner.

Here we present an example revealing how to design the filter banks so that the NB spectrogram can be correctly aligned with the low-frequency region of the WB spectrogram. For a given NB speech sampled at 8k Hz, the Mel-spectrogram represents a bandwidth only from 0--4k Hz, and the remaining 4k--8k Hz information is missing compared with the 16k Hz sampled WB data. According to the general formula for converting from Hertz to Mel scale frequency~\cite{deller2000discrete}, the associated relationship between the number of NB and WB filters is computed as follows:

\newcommand{\floor}[1]{\lfloor #1 \rfloor}

\begin{equation}\footnotesize
\label{equ:1}
M_N = \floor{M_W \times  \frac{\log (1 + f_N / 700)}{\log ( 1 + f_W / 700)}},
\end{equation}

where $f_N$ refers to the NB spectrogram upper limit, $f_W$ represents the WB spectrogram upper limit, and $M_N$ and $M_W$ denote the number of designed filters for the NB and WB data, respectively. In our implementation, we have $f_W=8000$, $f_N=4000$, and $M_W=64$. Therefore, according to Equation~(\ref{equ:1}), we obtain $M_N=48.3$. The feature dimension must be an integer, so we force the $M_N$ to 48 and set $f_N$ to 3978.69 Hz. In other words, the 3979.69 Hz to 4000 Hz information of the NB data is ignored.

\subsection{2-D CNN architecture}

Regarding the Mel-spectrogram as a visual image, we utilize the 2-D convolutions to learn the local time--frequency coupled patterns. For a given feature matrix of size $D \times T$, the 2-D convolution and pooling strategy brings us flexibility: the size of the input image, including the width and height, can be arbitrary. In speech processing, the 2-D CNN not only handles features with variable-length $T$, but also potentially accepts features with different dimensions $D$.

The representation learned by the 2-D CNN is a 3-D tensor of size $C\times H\times W$, where $C$ refers the number of channels, and $H$ and $W$ denote the height and width of the learned feature maps, respectively. We add a global statistics pooling~(GSP) layer after the 3-D feature maps to accumulate the global statistics over the time--frequency axes. We summarize a specific 2-D feature map $ {\bm{F}} \in \mathbb{R}^{H\times W}$ with a global mean and standard deviation statistics $\mu, \sigma$. Since there are $C$ channels of feature maps, we finally get a $2C$-dimensional vector $ {\bm{V}}={\left[\mu_1, \mu_2, \cdots, \mu_C, \sigma_1, \sigma_2, \cdots, \sigma_C\right]}$ to represent an arbitrary duration speech with different bandwidths.

\subsection{Training strategy}

The Mel-spectrogram trained with a 2-D CNN forms a new framework to potentially solve the mixed-bandwidth data problem. The remaining question is how to train a network that fits both the WB and NB spectrograms. Considering different scenarios, we investigate three kinds of training strategies, as described below.

\subsubsection{\textbf{Only WB data are given}}
\label{sec:only}

In this scenario, the evaluation dataset comprises both the WB and NB speech, but only WB speech is provided for training. Figure~\ref{fig:plot1} illustrates the proposed MB system training procedure. Here, WB spectrograms (64 dimensions) are reused to generate the NB spectrogram by selecting its sub-spectrograms (48 dimensions here). There are two rounds of parameter updates for a mini-batch of training data. The first update is from the full-size WB spectrogram, and the second update is from the sub-image of the WB spectrogram. It is desired that the network fits well on both WB and NB data, and the network parameters are shared for these two groups of images with different sizes. After the network is trained, we can feed it with either the full-size image (WB spectrogram) or the sub-image (NB spectrogram). The whole pipeline does not require any downsampling, upsampling, extension, or padding operation.

\subsubsection{\textbf{MB data from the same domain}}
\label{sec:samedomain}

Different from the situation in section~\ref{sec:only} where the NB training is simulated by selecting the sub-image from the WB spectrogram, here we have mixed training data with WB speech as well as NB speech. Therefore, we can first extract 64-dimensional WB spectrograms from the WB data and 48-dimensional NB spectrograms from both the NB and WB data; then, we can train the network as described in section~\ref{sec:only}. As illustrated in Fig.~\ref{fig:plot2}, the network parameters are shared across the WB and NB spectrograms; however, the output units representing the speaker identities are separate assuming there is no speaker overlapping between the WB and NB data. Although the network accepts features with different dimensions, the feature size within a mini-batch should be consistent. Therefore, we maintain two separated data loaders for the data with different bandwidths, and the mini-batch training data are fetched from these two data loaders alternatively to train the network.

\subsubsection{\textbf{MB data from different domains}}
\label{sec:diffdomain}

According to section~\ref{sec:samedomain}, the network parameters are shared across the training data of different bandwidths. In real-world applications, the MB and WB training data may be collected from different domains. Here we give a simple solution as illustrated in Fig.~\ref{fig:plot3}.  Our network consists of shared layers and multiple branches of domain-specific layers. Specifically, the bottom CNN is shared for both the NB and WB spectrograms to learn general feature representations. After the GSP layer, the fully-connected layer is learned independently for each domain. Therefore, the speaker embeddings and output units for different bandwidths are in separate branches. After the network is trained, the speaker embedding is extracted from the associated branch corresponding to the sampling rate.

\section{Experimental Results}

\subsection{Datasets}

\subsubsection{VoxCeleb1}

We first conduct simulated experiments on the VoxCeleb1 dataset~\cite{Chung:2018bp}. The training set includes \num{148642} utterances from \num{1211} celebrities. The test set contains \num{4715} utterances from the other \num{40} celebrities. The equal error rate~(EER) is used to measure the system performance.

At the beginning, both the training and test datasets were sampled at 16k Hz.  We obtained the 8k Hz evaluation data by downsampling the 16k Hz data using the Sox toolkit.

\newcommand{\blocka}[2]{\multirow{2}{*}{\(\left[\begin{array}{c}\text{3$\times$3, #1}\\[-.1em] \text{3$\times$3, #1} \end{array}\right]\)$\times$#2}
}
\newcommand{\blockb}[3]{\multirow{2}{*}{\(\left[\begin{array}{c}\text{3$\times$3, #2}\\[-.1em] \text{3$\times$3, #1}\end{array}\right]\)$\times$#3}
}
\renewcommand\arraystretch{1.1}
\setlength{\tabcolsep}{3pt}
\begin{table}[t]
    \caption{ The proposed network architecture. N/A: Not applicable}
        \centering
    \begin{adjustbox}{max width=0.99    \columnwidth}
        \begin{tabular}{|c|c|c|c|c|c|c|c|c|c|}
            \hline
            \textbf{Layer} &\textbf{Output size} & \textbf{Structure}&\textbf{ \#Params} \\
            \hline
            Conv1 & $\!16\!\times\!D\!\times\!T$& $3\!\times\!3$, stride 1& \num{176}    \\
            \hline
            \multirow{2}{*}{Res1}  &  \multirow{2}{*}{$16\!\times\!D\!\times\!T$} &  \blockb{16}{16}{3}\multirow{2}{*}{, stride 1}    &    \multirow{2}{*}{ 14K } \\
            &&&\\
            \hline
            \multirow{2}{*}{Res2}  &  \multirow{2}{*}{$32\!\times\!\frac{D}{2}\!\times\!\frac{T}{2}$} &    \blockb{32}{32}{4}\multirow{2}{*}{, stride 2} &\multirow{2}{*}{70K}  \\
            &  & &\\\hline
            \multirow{2}{*}{Res3}  &  \multirow{2}{*}{$64\!\times\!\frac{D}{4}\!\times\!\frac{T}{4}$} &   \blockb{64}{64}{6}\multirow{2}{*}{, stride 2}  &    \multirow{2}{*}{427K}     \\
            & & &\\
            \hline
            \multirow{2}{*}{Res4}  &  \multirow{2}{*}{$128\!\times\!\frac{D}{8}\times\!\frac{T}{8}$} &   \blockb{128}{128}{3}\multirow{2}{*}{, stride 2}    &    \multirow{2}{*}{821K} \\
             & & &\\\hline

            GSP&256& Statistics pooling  &0\\

            \hline
            FC1 (embedding) &$ 128$ &Fully-connected & 32K\\
            \hline
            FC2 (Output) & speakers &Fully-connected &   N/A   \\
            \hline
        \end{tabular}
    \end{adjustbox}
    \label{tab:resnetconfigt}
       \vspace{-0.4cm}
\end{table}

\begin{table} [t]
    \caption{  Performance on the VoxCeleb1 test data when only the WB VoxCeleb1 training data are given. $\rightarrow$: sub-spectrogram selection operation. SR: Sampling rate.}
    \centerline {
      \resizebox{0.98\columnwidth}{!}{
        \begin{tabular}{c c c c c  c c c c}
        \toprule
         \textbf{ID}&\textbf{Method}& \textbf{Training SR and \#Filters} & \textbf{Testing SR}&\textbf{Testing \#Filters}&\textbf{EER (\%)}\\
        \midrule
        1&\multirow{4}{*}{  WB baseline} &\multirow{4}{*}{16k and  64}& 16k &  64 &\textbf{4.35}&\\
          2&    &&8k&64&20.13&\\
            3&   &&8k&48&8.82&\\
              4&  &&16k&64$\rightarrow$48&8.87&\\
         \cmidrule(lr){1-6}
            5&              \multirow{4}{*}{ NB baseline} &\multirow{4}{*}{8k and 48}& 8k &  48 &\textbf{4.92}&\\
              6&&&16k&48&18.86\\
               7&&&16k&64&8.01&\\
                8&&&16k&64$\rightarrow$48&4.95&\\
                         \cmidrule(lr){1-6}
                   \multirow{2}{*}{  \textbf{9}}&     \multirow{2}{*}{  \textbf{Proposed MB}}& \multirow{2}{*}{16k and 64\&48}& 16k&64&\textbf{4.07}\\
                        &&&8k&48&\textbf{4.37}\\
                        \bottomrule
    \end{tabular}}}
    \label{table:onlywb}
    \vspace{-0.3cm}
\end{table}

\subsubsection{SITW and NIST SRE 2016}

The SITW dataset consists of unconstrained audio--visual data of English speakers~\cite{mclaren2016speakers}. Our focus is on its core--core protocol for both the development and evaluation sets. For the NIST SRE 2016, the test data are composed of telephone conversations collected outside North America, spoken in Tagalog and Cantonese~\cite{sadjadi20172016}. The development set contains some unlabeled data, which is useful for the unsupervised domain adaptation.

The pooled VoxCeleb1 and VoxCeleb2 datasets were used as our training set for the evaluation on SITW. Finally, a training set of \num{1236567} utterances from \num{7185} celebrities was obtained. We refer to these data as VoxCeleb1\&2 16k data. We also have NB training data from NIST SRE 2004--2010, Mixer 6, Switchboard 2 Phase 1, 2, and 3, as well as Switchboard Cellular. There is a total of \num{99661} utterances from \num{7222} speakers. We refer to this data as SRE 8k data.

\subsection{Implementation details}

First, 64- and 48-dimensional Mel spectrograms are extracted for the WB and NB data, respectively.
Our network is based on the ResNet~\cite{He2016Deep} structure, and the architecture is described in Table~\ref{tab:resnetconfigt}. Dropout with a rate of 0.5 is added before the softmax layer, and the network is trained with a typical cross-entropy loss. We adopt the common stochastic gradient descent algorithm with momentum \num{0.9} and weight decay 1e-4.

\begin{table*} [t]
    \caption{  Performance on the SITW and NIST SRE 2016 datasets. }
    \centerline {
        \begin{tabular}{c c c c c  c c c c}
        \toprule
         \multirow{3}{*}{\textbf{ID}}& \multirow{3}{*}{\textbf{Training data}}& \multicolumn{5}{c}{\textbf{Testing  \textbf{EER (\%)} }}\\
           \cmidrule(lr){3-7}
       & &\multicolumn{2}{c}{\textbf{SITW }} & \multicolumn{3}{c}{\textbf{NIST SRE 2016 }}\\
            \cmidrule(lr){3-4}     \cmidrule(lr){5-7}
        &    &\textbf{Dev}&\textbf{Eval}&\textbf{Pool}&\textbf{Cantonese}&\textbf{Taglog}\\
        \midrule
        1&VoxCeleb1\&2 16k&\textbf{2.17}& \textbf{2.49}& 16.88 & 10.86  &23.02 \\
          2& SRE 8k&14.29&17.52&\textbf{6.19}&\textbf{3.66} &\textbf{8.61} \\
          \midrule
          3&    VoxCeleb1\&2 8k +  SRE 8k&3.20&3.52&5.69 &  3.39  &8.10 \\
           4& \textbf{VoxCeleb1\&2 16k + SRE 8k}&\textbf{2.91}&\textbf{3.18}&\textbf{5.44}&  \textbf{3.10} & \textbf{7.61}\\
                        \bottomrule
    \end{tabular}}
    \label{table:mixdiffdomain}
       \vspace{-0.3cm}
\end{table*}

\begin{table} [t]
    \caption{ Performance on the VoxCeleb1 test data when the VoxCeleb1 training data are split into two subsets.}
    \centerline {
        \begin{tabular}{c c c c c  c c c c c c}
        \toprule
         \multirow{3}{*}{\textbf{ID}} &  \multirow{3}{*}{\textbf{VoxCeleb1 Training data}}&  \multicolumn{4}{c}{\textbf{VoxCeleb1 Testing}}\\
           \cmidrule(lr){3-6}
        &&\multicolumn{2}{c}{\textbf{\# filters}}&\multicolumn{2}{c}{\textbf{EER (\%)}}\\
        \cmidrule(lr){3-4} \cmidrule(lr){5-6}
        &&\textbf{16k}&\textbf{8k}&\textbf{16k}&\textbf{8k}\\
        \midrule
        1&Subset1 16k&64&48&\textbf{5.66}&11.08\\
        2&Subset2 8k&64$\rightarrow$48&48&6.11&\textbf{6.07}\\
          \midrule
          3&    Subset1 8k + Subset2 8k &64$\rightarrow$48&48&4.95&4.92\\
            4&\textbf{Subset1 16k + Subset2 8k}&64&48&\textbf{4.53}&\textbf{4.51}\\
                        \bottomrule
    \end{tabular}}
    \label{table:mixsamedomain}
       \vspace{-0.3cm}
\end{table}

In total, there are three downsampling operations within the convolutional layers. Therefore, the original Mel-spectrograms are downsampled to compact feature maps with size $\frac{D}{8} \times \frac{T}{8}$ before the GSP layer. In the training phase, we implement a variable-length data loader to generate mini-batch training samples on the fly~\cite{caiwch_taslp}. For each step, a dynamic mini-batch of data with size $B \times D \times T$ is generated, where $B$ is the mini-batch size, $D$ is the feature dimension, and $T$ is a batch-wise variable frame length ranging from \num{300} to \num{800}.

After the network is trained, the 128-dimensional speaker embeddings are extracted for evaluation. Simple cosine similarity is adopted to compute the pairwise score for the VoxCeleb1 and SITW evaluation trials. For the NIST SRE 2016 experiments, we adopt the adaptive PLDA backend as implemented in the Kaldi SRE16 recipe~\cite{SRE16v2}.

\subsection{Results and discussion}

\subsubsection{\textbf{Only WB data are given}}

We first train a WB baseline system using the 16k training data. IDs from 1 to 4 in Table~\ref{table:onlywb} show the results for different setups of the evaluation data. It reveals that if we feed the WB model with NB data, then the 48-dimensional NB spectrogram obtains much better results than the 64-dimensional one~(EER 8.87$\%$ vs  20.13$\%$). After selecting the 48-dimensional sub-image from the 64-dimensional WB spectrogram, the system obtains almost the same result as the 8k NB test data. These results suggest that it is more crucial to keep the resolution of the features consistent rather than the dimension the features. We also train an NB system using the downsampled 8k data. We reach similar conclusions as for the WB system, but the best NB results is an EER of 4.92$\%$, which is slightly worse than that of the WB system (4.35\%). This indicates that the WB features might contain more useful information than the NB ones.

By applying corresponding models for the test data with each sampling rate separately, the systems achieve 4.35\% EER for the WB test data and 4.92\% for the NB test data. Following the approach described in section~\ref{sec:only}, we train the proposed MB system using only the 16k WB data. Our single MB model achieves 4.07\% and 4.37\% EER for the WB and NB evaluation data, respectively.

\subsubsection{\textbf{MB data from the same domain}}

Here, the VoxCeleb1 training data are divided into two subsets: the first part consists of \num{74438} utterances from \num{622} speakers, and the second part includes \num{74204} utterances from \num{589} speakers. All of the speech data in the second part are downsampled to 8k Hz to simulate the NB training data. The top half of Table~\ref{table:mixsamedomain} shows the results of the single WB/NB system trained by only subset1 16k or subset2 8k data with the best choices of testing spectrograms. Compared with the results in Table~\ref{table:onlywb}, the performance here is degraded since the scale of the training data is reduced.

To utilize the training data with different bandwidths, we first train a baseline system by downsampling the data from the first part to 8k Hz and then pooling together the data from the two subsets. We can see that the pooled 8k system (ID 3) achieves much better results than the system trained by any single subset (ID 1 and 2). Following the approach described in section~\ref{sec:samedomain}, we train the proposed MB system (ID 4) by jointly modeling the original subset1 16k and subset2 8k data. Compared with the baseline system (ID 3), we obtain consistent EER reduction for both the WB and NB evaluation data.

\subsubsection{\textbf{MB data from different domains}}

The top half of Table~\ref{table:mixdiffdomain} shows the results of the WB system trained with the VoxCeleb1\&2 16k data and the NB system trained with the SRE 8k data. These two systems don't work well on each other's evaluation sets, possibly due to the additional domain mismatch. We further develop a pooled NB system by downsampling the VoxCeleb1\&2 16k data to 8k Hz and pooling the two training datasets in system ID 3. Compared with the SRE 8k system (ID 2), the pooled NB system performs much better on NIST SRE 2016 since the number of training utterances and speakers are increased. However, on the SITW dataset, the performance is degraded significantly compared with the VoxCeleb1\&2 16k WB baseline system (ID 1), which might be due to the domain mismatch.
Following the approach described in section~\ref{sec:diffdomain}, we train an MB system (ID 4) by jointly modeling the VoxCeleb1\&2 16k and SRE 8k data in a single network. The proposed MB system consistently outperforms this baseline (ID 3) on both SITW and NIST SRE 2016 evaluation sets.

\section{Conclusion and future work}

In this paper, we propose a novel multi-bandwidth joint modeling approach for speaker verification. We show that speech data with different sampling rates can be flexibly integrated in a single speaker embedding model based on a 2-D CNN without any additional downsampling, upsampling, extension, or padding operations. Experimental results show that the proposed MB systems achieve significant improvement on both the NB and WB evaluation data. Our future works include comparing the proposed systems with other state-of-the-art MB solutions, such as using a DNN-based bandwidth extension module as the frontend for all NB data.

\section{Acknowledgement}
This research is funded in part by the National Natural Science Foundation of China (61773413), Key Research and Development Program of Jiangsu Province (BE2019054), Six talent peaks project in Jiangsu Province (JY-074), Science and Technology Program of Guangzhou City (201903010040,202007030011).

\newpage
\bibliographystyle{IEEEbib}

\bibliography{caiwch_taslp}

\end{document}